# Unified Theory of Activated Relaxation in Liquids over 14 Decades in Time

Stephen Mirigian and Kenneth S. Schweizer*

Departments of Materials Science & Chemistry, University of Illinois, Urbana, IL 61801

*kschweiz@illinois.edu

## Abstract

We formulate a predictive theory at the level of forces of activated relaxation in hard sphere fluids and thermal liquids that covers in a unified manner the apparent Arrhenius, crossover and deeply supercooled regimes. The alpha relaxation event involves coupled cage-scale hopping and a long range collective elastic distortion of the surrounding liquid, which results in two inter-related, but distinct, barriers. The strongly temperature and density dependent collective barrier is associated with a growing length scale, the shear modulus and density fluctuations. Thermal liquids are mapped to an effective hard sphere fluid based on matching long wavelength density fluctuation amplitudes, resulting in a zeroth order quasi-universal description. The theory is devoid of fit parameters, has no divergences at finite temperature nor below jamming, and captures the key features of the alpha time of molecular liquids from picoseconds to hundreds of seconds.

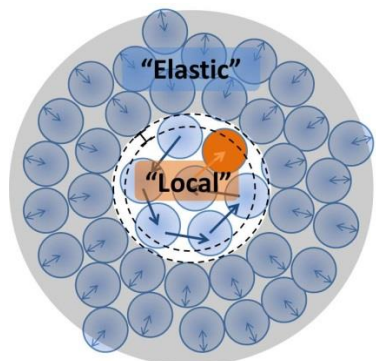

The cooling of glass-forming liquids leads to a spectacular increase in the structural relaxation time and viscosity by 14 or more orders of magnitude, traditionally parsed into physically distinct regimes separated by material-specific "crossover temperatures"[1-3]. The origin of this remarkable behavior remains vigorously debated, with the many conceptually different theories[4,5] often crafted to treat only one regime. Confrontation with experiment generally involves empirically-adjusted fit parameters which renders a definitive evaluation of underlying ideas difficult. Recent simulations suggest the few existing "microscopic" theories have major limitations[6].

Ideal mode coupling theory[7] (IMCT) is a first principles force-level approach which successfully captures the onset of transient localization. Unfortunately, it predicts literal kinetic arrest far above $T_g$ due to the neglect of activated processes and (at best[8]) is relevant to just a few orders of magnitude of the initial slowing down. The nonlinear Langevin equation (NLE) approach[9] builds on the microscopic MCT framework to treat *single* particle hopping on the cage scale (Fig.1), and for hard spheres there are no divergences below jamming. Activated relaxation in the precursor regime and multiple nongaussian dynamically heterogeneous phenomena in concentrated hard sphere and colloidal fluids seem well described[10]. However, longer range collective motions are not captured, a presumably fatal limitation in ultra-dense and deeply supercooled liquids[6,11].

In this Letter we formulate a unified, no fit parameter, force-level theory at the level of molecules that can describe relaxation in liquids from ~psec to 100 secs based *solely* on activated dynamics. The latter perspective is inspired by the view of Kivelson and Tarjus[12,13] (and others) where distinct "molecular" and "cooperative" barriers dominate at high and low temperatures, respectively. Normal liquids do display an *apparent* Arrhenius (barrier, $E_A$), which is not understood[1-3,12-17]. Experiments[14,16-18] find striking correlations between $E_A$, $T_g$, and fragility, suggesting a connection between the "uncooperative" activated and "cooperative" super-Arrhenius relaxations. Our goal is to address both activated regimes within a single framework.

Our starting point is NLE theory as formulated for the *angularly-averaged* displacement of a tagged sphere (diameter $d$) from its initial position, $r(t)$, which obeys the evolution equation[9]: $\zeta_s dr / dt = -\partial F_{dyn}(r;\phi) / \partial r + R(t)$, where $\zeta_s$ is the short time friction constant and R(t) the corresponding random force. The key quantity is a *displacement-dependent* dynamic free energy, $F_{dyn}(r)$, which quantifies the effective force exerted on a particle due to its surroundings[9]

$$\beta F_{dyn}(r;\phi) = -3ln(r/d) - \frac{3\phi}{\pi^3} \int_0^\infty dq q^2 \frac{C^2(q)S(q)}{1+S^{-1}(q)} e^{-\frac{(qr)^2}{6}\left[1+S^{-1}(q)\right]}$$ (1)

where $\beta \equiv 1/k_B T$, $\phi \equiv \pi\rho d^3 / 6$ is the fluid volume fraction, $S(q) = (1-\rho C(q))^{-1}$ is the structure factor and $C(q)$ the direct correlation function. For hard spheres, beyond $\phi_c \cong 0.43$ (using Percus-Yevick theory for structural input) $F_{dyn}(r)$ acquires a minimum (transient localized state) at $r_{loc}$ and a barrier at $r_B$ of height $F_B$ indicating the emergence of hopping transport with a mean jump length $\Delta r = r_B - r_{loc}$ (Fig. 1); both $F_B$ and $r_B$ grow with $\phi$ while $r_{loc}$ shrinks.[9,10] Analysis of the incoherent dynamic structure factor at the cage peak, $q^* \approx 2\pi / d$, yields a mean activated alpha relaxation time of (see SI):

$$\tau_\alpha = \tau_s \frac{\pi}{(q^* d)^2} \sqrt{\frac{K_0}{K_B}} e^{\beta F_B}$$

(2)

where $\tau_s = g_d^2 \tau_E$, $\tau_E^{-1}$ is the binary collision rate, $g_d$ the contact value of g(r), and $K_0$ ( $K_B$ ) the absolute magnitude of the curvature of $F_{dyn}$ at the well (barrier). The glassy dynamic plateau shear modulus associated with the transiently localized state is[7,9]

$$G' = \frac{k_B T}{60\pi^2} \int_0^\infty dq \left( q^2 \frac{1}{S(q)} \frac{\partial S(q)}{\partial q} \right)^2 e^{-\frac{(qr_{loc})^2}{3S(q)}} \qquad (3)$$

Very importantly for the present work, NLE theory predicts (at sufficiently high $F_B$) multiple connections between short and long time properties due to the microscopic foundation of $F_{dyn}$ and its Einstein glass description[9,19] of the localized state[20]

$$\frac{G'd^3}{k_B T} \frac{5\pi}{3\phi} \quad = \quad \beta d^2 K_0 \quad = \quad \frac{3d^2}{r_{loc}^2} \quad \propto \left( \beta F_B \right)^2 . \qquad (4)$$

The q-dependent "dynamical vertex" in Eq(1) is dominated by[20] *high* wavevector force correlations corresponding in real space to distances less than the cage radius, $r_{cage}(\phi) \approx 3d/2$. The high wavevector dominance is a generic feature of MCT and NLE theory due primarily to factorization of multi-point correlations. Recent work has shown this local character results in too weak a growth of the alpha time in the ultra-dense regime of hard sphere fluids[6,11], and below we explicitly document this failure for thermal liquids. For activated motion, such a breakdown of NLE theory is perhaps not unexpected since hopping is a large amplitude event[9,10], $\Delta r \rightarrow (0.3 - 0.4)d$, and at high $\phi$ there is not enough room for this motion to uncooperatively occur.

Our central strategy for going beyond NLE theory adopts an elastic perspective[21] of collective motion. Specifically, in cold/dense liquids, steric considerations require increasing participation by the surrounding medium in the form of a small cage expansion and long range elastic displacements of many particles in order to allow local hopping. The alpha process is thus of a mixed local-nonlocal character (as sketched in Fig.1): a core of hopping particles on the $2r_{cage} \approx 3d$ scale "dressed" by an elastic distortion. This real space picture is supported by glassy colloid experiments that find the elementary irreversible relaxation event consists of a core of particles of radius of $\approx 1.5d$ with hop amplitudes $\approx (0.5 \pm 0.25)d$ surrounded by an elastic field[22], which is also consistent with NMR measurements on deeply supercooled molecular liquids[23].

We qualitatively extend elastic models[21] to include the local activated process, its microscopic coupling to the collective process, and a growing length scale. We first develop the ideas for hard spheres. To calculate the collective barrier, recall that in NLE theory each caged particle moves *isotropically* in a dynamic mean field $F_{dyn}(r)$ (see Fig. 1) and hops in a random direction a mean distance $\Delta r = r_B - r_{loc}$. To be consistent with the angularly-averaged $F_{dyn}(r)$, we assume cage expansion is spherically symmetric. A straightforward geometry calculation yields a *mean* outward radial cage expansion of

$$\Delta r_{eff} = \frac{3}{r_{cage}^3} \left[ \left( \frac{\Delta r}{4} - r_{cage} \right) \left( r_{cage}^3 - \left( r_{cage} - \frac{\Delta r}{4} \right)^3 \right) + \frac{1}{4} \left( r_{cage}^4 - \left( r_{cage} - \frac{\Delta r}{4} \right)^4 \right) \right] \quad (5)$$

Since $\Delta r(\phi) \le 0.4d$, $\Delta r/4 \ll r_{cage}$ and thus Eq(5) simplifies to $\Delta r_{eff} \to 3\Delta r^2 / 32 r_{cage}$, which is orders of magnitude smaller than the jump distance and also less than $r_{loc}$[9,20]. This justifies using linear elasticity to compute the displacement field[24,25]: $u(r) = \Delta r_{eff}(\phi) \left[ r_{cage} / r \right]^2$, $r \ge r_{cage}$. The collective elastic barrier then follows by summing the harmonic displacements outside the cage:

$$F_{elastic} = 4\pi\rho \int_{r_{cage}}^{\infty} dr\, r^2 g(r) \left[ \frac{1}{2} \frac{\partial^2 F_{dyn}(r)}{\partial r^2}\bigg|_{r_{loc}} u^2(r) \right] = 12\phi \Delta r_{eff}^2 \left( \frac{r_{cage}}{d} \right)^3 K_0 \qquad (6)$$

where $g(r) \approx 1$. Since the integrand decays as $r^{-2}$, capturing 90% of $F_{elastic}$ requires including $\phi$–*dependent* displacements to $r \approx 10 r_{cage} \approx 15d$, and in this sense is long range. We emphasize that $F_{elastic}$ scales with the spring constant, $K_0$, that quantifies transient localization in the Einstein model spirit that defines NLE theory.

The total barrier in Eq(2) is taken to be additive, $F_{tot} \equiv F_B(\phi) + F_{elastic}(\phi)$, since we view the activation process as a single event where a local hop requires an expansive cage fluctuation to locally create "free volume". The two barriers are *not* independent and depend on the *same* equilibrium input. Their growth with increasing $\phi$ or reduced alpha time is shown in the lower inset of Fig. 2. The local barrier initially dominates consistent with NLE theory being useful in the precursor regime. But the more rapidly increasing collective contribution wins in the strongly viscous regime. Note that the rate of change of the two barriers with logarithmic relaxation time or $\phi$ are roughly equal at $\tau_\alpha \approx 10^4 \tau_s$ ($\approx 10^{-8} s$ for thermal liquids) defining a "crossover" barrier of $F_{tot} \approx 10 k_B T$ .

Based on the above radially symmetric displacement field, Dyre has proposed a phenomenological shoving model for deeply supercooled liquids where the dominant barrier scales with the shear modulus, $F_{shove} = G'V_c$ , where $V_c$ is an empirical *constant* "characteristic volume" deduced by fitting data[21,25]; Dyre has also shown that corrections to $F_{shove}$ due to possible anisotropy of the displacement field are small[26]. Using Eq(4), Eq(6) can be rewritten as: $F_{elastic} = 12\phi \Delta r_{eff}^2 r_{cage}^3 d^{-3} K_0 \equiv G'V_c$ , where a *microscopic* cooperative volume is identified as $V_c = 20\pi\Delta r_{eff}^2 r_{cage}^3 d^{-2}$. Note we have not assumed G'

determines the barrier, but rather this follows as a consequence of Eqs (4) and (6). Figure

2 shows $V_c$ is significantly smaller than the molecular volume, $V_m \equiv \pi d^3/6$; converted

to a length scale, at vitrification $\xi_s \equiv (6V_c/\pi)^{1/3} \approx 0.3d$ corresponding to

$\xi_s \approx 0.2 - 0.35\,nm$ for typical molecules, consistent with experimental fits[21,25,27].

However, $V_c$, and hence the displacement field amplitude, grows significantly with $\tau_\alpha$

via the cage expansion factor $\Delta r_{eff}^2(\phi)$. Moreover, the degree (as quantified by logarithmic derivatives in the inset) to which $V_c$ and the shear modulus determine the growth rate of $F_{elastic}$ are nearly equal up to $F_{tot} \approx 10k_BT$, beyond which G' becomes increasingly more important. Despite this more complex origin of $F_{elastic}$, plus a local barrier that increases with $\phi$, we empirically find (Fig.2) $\log(\tau_\alpha) \propto G'$ is well obeyed over ~8-10 orders of magnitude of slowest relaxation, consistent with experiments[21,25,27].

A key conceptual feature is the collective barrier arises from *both* a growing

length scale, $l \propto \Delta r_{eff}(\phi)$, and emergent shear rigidity. This connects to a recent

speculation[8] that the shoving and entropy crisis[19] ideas might be (partially) reconciled if

$V_c$ grew with cooling or densification; our results also provide microscopic insight as to

why[8] $\xi_s$ is so small. In the strongly viscous regime we numerically find $\beta F_{elastic} \propto (\beta F_B)^2$

,which to leading order can be analytically understood from Eqs.(4) and (6) as a consequence of the common, but different, scaling of the barriers with $r_{loc}$. The two barriers are thus related, which has significant consequences.

The theory will be applied to simulations and experiments on hard sphere fluids in future work. Here, since only 3-5 orders of magnitude of slowing down in the precursor regime are probed in this system, as a far more demanding test of our ideas we study thermal liquids including the deeply supercooled regime.

Motivated by our dual goals of being able to generally treat structurally complex molecular liquids and to construct a zeroth order quasi-universal description[13], we propose a new mapping to a reference hard sphere (HS) fluid. Given the structural underpinnings of our approach, the *molecular* level long wavelength dimensionless density fluctuation amplitude, $S(q \to 0) \equiv S_0$, of the thermal and reference HS liquids are required to be equal. This defines a material and temperature dependent HS volume fraction via $S_{0,mol}^{expt}(T) = \rho k_B T \kappa_T \equiv S_0^{HS}(\phi_{eff}(T))$ where $\rho$ is the molecular number density and $\kappa_T$ the isothermal compressibility. Solely to allow an interpretation of how $S_0$ is related to chemical variables and molecular mass or size, we define a "site" level analog of the mapping in terms of $S_{0,site}^{expt}(T) = S_{0,mol}^{expt} N_s$ where rigid molecules consist of $N_s$ sites and a united atom model of CH, CH$_2$, CH$_3$, OH groups is adopted. We consider the widely studied molecular liquids ortho-terphenyl (OTP, C$_{18}$H$_{18}$, $N_s = 18$), tris-napthylbenzene (TNB, C$_{36}$H$_{36}$, $N_s = 36$) and glycerol (G, $N_s = 6$). Prior analysis[28] has shown experimental dimensionless compressibility data is well described by $\left( S_{0,site}^{expt}(T) \right)^{1/2} = -A + \left( B/T \right)$, where B scales as the cohesive energy and A is an entropic packing-related factor. The key energy scale is thus $\sqrt{N_s} B$, which implies for chemically homologous molecules (e.g., OTP, TNB) the characteristic dynamical temperatures will differ only due to $N_s$.

Using equation-of-state data at 1 atm (see SI) to compute $S_0^{HS}(\phi_{eff}(T))$, Figure 3 presents no fit parameter calculations and comparison to experiments for G[29], OTP[30] and TNB[31,32]. The Angell plot of the main frame shows multiple notable features. (i) The local NLE theory result for OTP follows an *apparent* Arrhenius behavior. This agrees with the full theory at high temperatures over the range $\tau_\alpha \sim 1$-100 psec, and also experimental observations[2,3]. However, the NLE approach clearly breaks down in the deeply supercooled regime. (ii) The value of $E_A$, in absolute terms or relative to $T_g$, is sensitive to how the calculations are fit and also high temperature details of the theory. However, our estimates for OTP and other van der Waals molecular liquids are within a factor of 2 of experiment[2,3,14,15]. The predicted emergence of an apparent Arrhenius behavior at a high $T_A \sim 2 T_g$ is also consistent with observations[2,3,14-16] and Frenkel line arguments[33]. What we predict for other classes of liquids (alcohols, metals, ionic, network formers) remains to be determined. (iii) Defining vitrification as when $\tau_\alpha = 100s$, the full theory yields a $T_g$ of 205 K, 271 K, and 366 K for G, OTP and TNB, respectively, within ~6-10% of the measured values of 191 K, 246 K, and 344 K. (iv) As anticipated based on our mapping, relaxation curves for OTP and TNB nearly collapse. (v) The dynamic fragility, $m = \partial \log(\tau_\alpha) / \partial (T_g / T)|_{T_g}$, is 54 (G), 74 (OTP) and 78 (TNB), compared to experimental values of 49, 81 and 86. The controlling parameter for the fragility can be shown to be proportional to $A\sqrt{N_s}$, which reflects the packing contribution to the temperature dependence of $S_0(T)$, and also controls the separation between the onset of activated dynamics at $T_A$ and $T_g$. The good quality of our predictions for H-bonding glycerol are surprising, and future work is required to determine whether it is robust or fortuitous.

The level of agreement of the no fit parameter theory with experiments over ~11-14 orders of magnitude of $\tau_\alpha$ variation is encouraging. Although in equilibrium the barriers and $V_c$ grow indefinitely with increasing $\phi$ or cooling, there are no divergences above zero Kelvin or below jamming[11,20]. As a distinct test of the theory and mapping, the inset of Fig. 3 presents calculations of the glassy shear modulus for G and OTP which probes the minimum of $F_{dyn}(r)$ per Eq(4). The glycerol data[27] demonstrates the growth of shear rigidity with cooling is accurately captured. As anticipated from Fig. 2, we find (not plotted) the shoving model result[21,25] $\log(\tau_\alpha / \tau_0) \propto X(T) \equiv T_g G'(T) / TG'(T_g)$ is well obeyed over the slowest 8-10 orders of magnitude, with upward deviations at higher temperatures (per experiment) primarily due to the local barrier.

We now analyze our calculations as "numerical data" in Arrhenius, MCT, dynamic facilitation, and entropy crisis frameworks; representative results are shown in Fig.4 for OTP. An apparent Arrhenius regime extends from $\tau_\alpha \approx 1-100 \, ps$ over $T_A$~558K~$2T_g$ to 380K, terminating at $T_{A,eff}$~380K~$1.4T_g$. Below $T_{A,eff}$, a narrow region can be empirically fit ala MCT, $\tau_\alpha \propto (T-T_c)^{-\nu}$, with $T_c$~$1.12T_g$, per experimental estimates[1,34,35] of $T_c \sim (1.14 \pm 0.02)T_g$; the critical power law form empirically works when the local and cooperative barriers are comparable, failing when $\tau_\alpha > 10^{-7} - 10^{-6} \, s$.

In the deeply supercooled regime, a "parabolic law" based on arrow kinetic constraint models (KCM)[36], $\log(\tau_\alpha/\tau_0) = (J/T_0)\left[(T_0/T)-1\right]^2$, fits diverse experimental data very well below an empirically-deduced onset temperature, $T_0$. Figure 4 demonstrates this formula describes our calculations extremely well over ~12 orders of magnitude, and the extracted parameters of $T_0 \approx 1.4T_g \approx T_{A,eff}$, $J/T_0 \approx 8$, $\tau_0 \approx 4\cdot10^{-10}s$ agree with the prior KCM fit of data[36]. In our theory, the parabolic form is primarily a consequence of the connection $\beta F_{elastic} \propto \left(\beta F_B\right)^2$. The inset of Fig. 4 applies the Stickel derivative analysis[35] which linearizes relaxation curves if the Vogel-Fulcher-Tamman (VFT) formula, $\ln\left(\tau_\alpha/\tau_0\right) \propto B/\left(T-T_{vft}\right)$, applies. Our calculations are rather well linearized per the empirical "high and low temperature VFT laws" seen experimentally[1-3,35]. Their intersection defines a crossover temperature $T_B \approx 1.19T_g$ at $\tau_\alpha \approx 5.6\cdot10^{-7}s$, values in good agreement with OTP experiments[2,3,34,35]; this crossover reflects the presence of two barriers with $T_B$ determined by when their thermal growth rates are roughly equal (Fig. 2). If one naively extrapolates the low temperature VFT line to $\tau_\alpha \to \infty$ we find $T_{vft} \approx 0.76T_g$, consistent with values deduced from data fitting[34,35].

In conclusion, we have formulated a force-level predictive theory of the alpha relaxation time that captures in a unified manner the normal, crossover and deeply supercooled regimes based on activated dynamics. The elementary relaxation event is of a mixed local-nonlocal character, involving distinct (but *not* independent) cage-scale and collective elastic barriers. Our mapping provides a zeroth order quasi-universal description in the sense that liquids are related to an effective HS fluid via the thermal density fluctuation amplitude. The effect of applied pressure, and connections between isobaric and isochoric relaxation, will be studied in future work. The new theory sets the stage for quantitatively treating colloidal suspensions, polymer liquids, thin films and other systems. It also provides a starting point for improving the theory in many directions including the explicit role of molecular shape and dynamic heterogeneity.

**Acknowledgement.** This work was supported by the U.S. Department of Energy, Basic Energy Sciences, Materials Science Division via Oak Ridge National Laboratory.

**Supporting Information Available**: Derivation of Eq(2) and explanation of the equation-of-state data used to perform the calculations in Figs 3 and 4. This material is available free of charge via the Internet at http://pubs.acs.org

# Figure Captions

**Figure 1**: Schematic of the conceptual elements of the theory: (a) coupled local hopping and cage expansion, (b) dynamic free energy with key length and energy scales indicated, (c) long range harmonic displacement of particles outside the cage.

**Figure 2:** Dimensionless alpha relaxation time (points) as a function of the dimensionless shear modulus; the dashed line indicates exponential behavior. Upper inset: Relative contribution, as quantified by logarithmic derivatives, to the collective barrier growth due to the cooperative volume and shear modulus. Lower inset: NLE (black dash-dotted), collective (green dotted), total (blue solid) barriers versus logarithmic alpha time (bottom) and volume fraction (top). Right axis shows the cooperative volume (red dashed line).

**Figure 3:** Alpha time versus dimensionless inverse temperature (solid curves) and corresponding data for glycerol[29] (blue circles), OTP[30] (red squares), TNB[31,32] (yellow diamonds). The dashed curve is the nearly Arrhenius NLE theory result for OTP. Inset: Glycerol shear modulus (in GPa) as a function of temperature: blue circles are experiment[27] (shifted up by a factor of 2.75), and smooth curve is the theoretical result using $d = 6.2\mathring{A}$. The theory result for OTP ($d = 9.1\mathring{A}$) is the lower red curve. The hard sphere diameters were determined from the experimental bulk liquid density and their weak temperature dependence has been ignored.

**Figure 4:** Fits of the theoretical relaxation time for OTP (blue circles) in different regimes to Arrhenius (blue straight line), critical power (red dashed curve), and parabolic (orange solid curve) laws. Inset: Stickel VFT analysis based on plotting $\left[ (d / d(1/T)) \ln(\tau_\alpha) \right]^{-1/2}$ vs. $T^{-1}$.

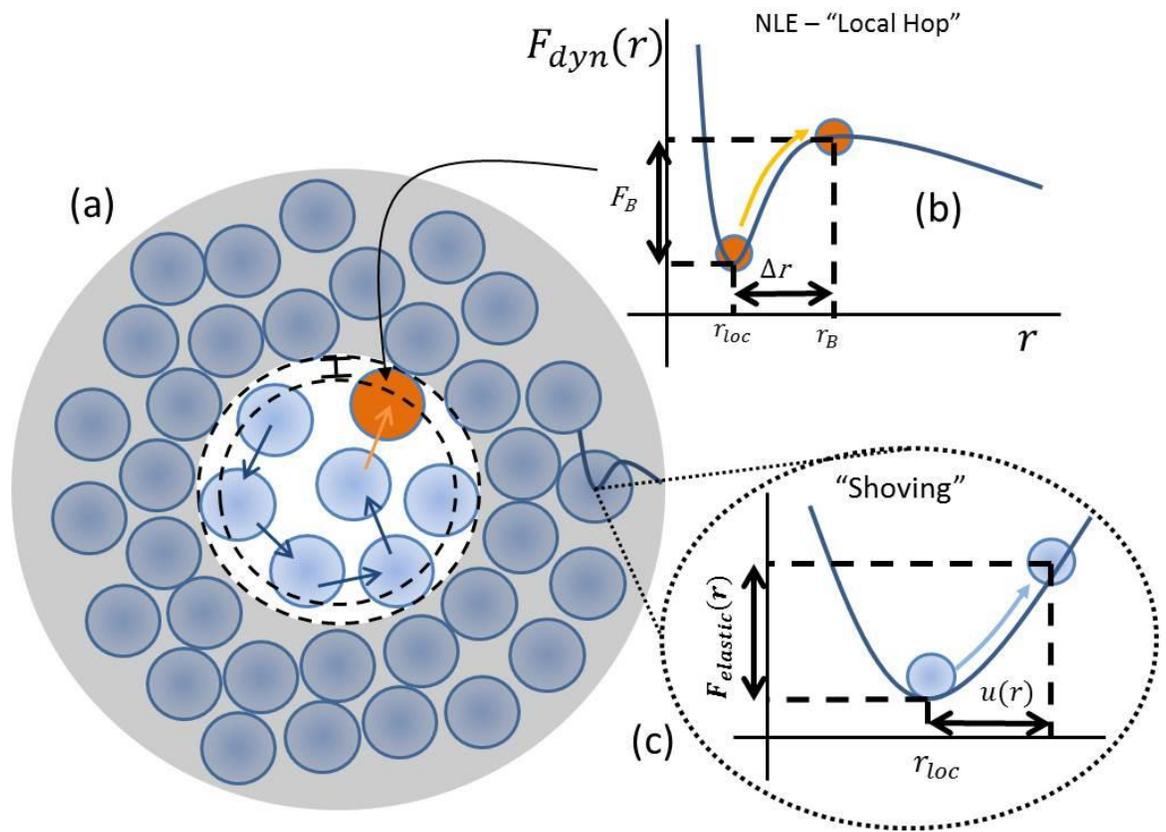

**Figure 1**

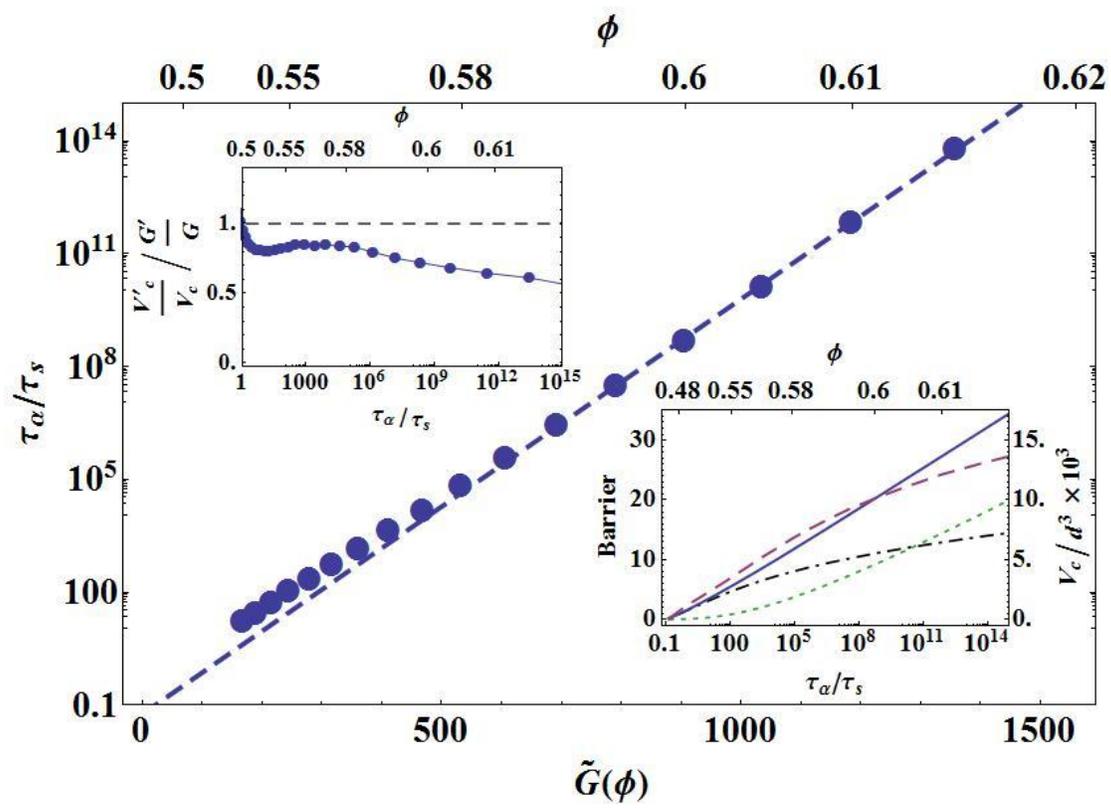

**Figure 2**

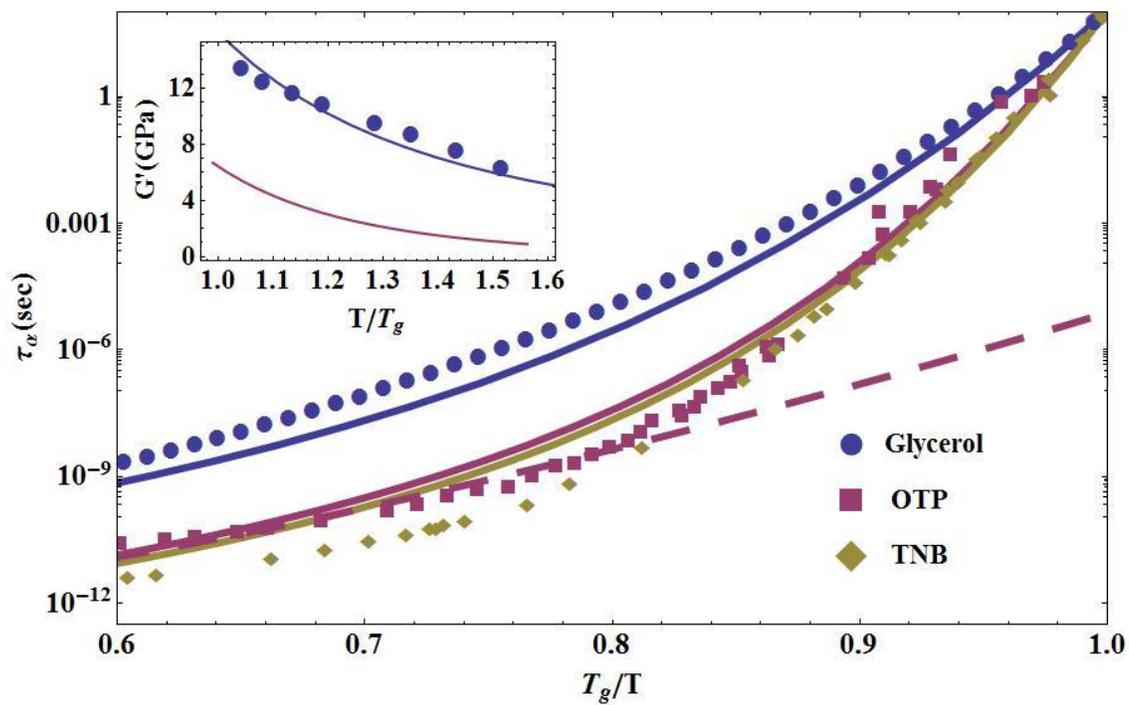

**Figure 3**

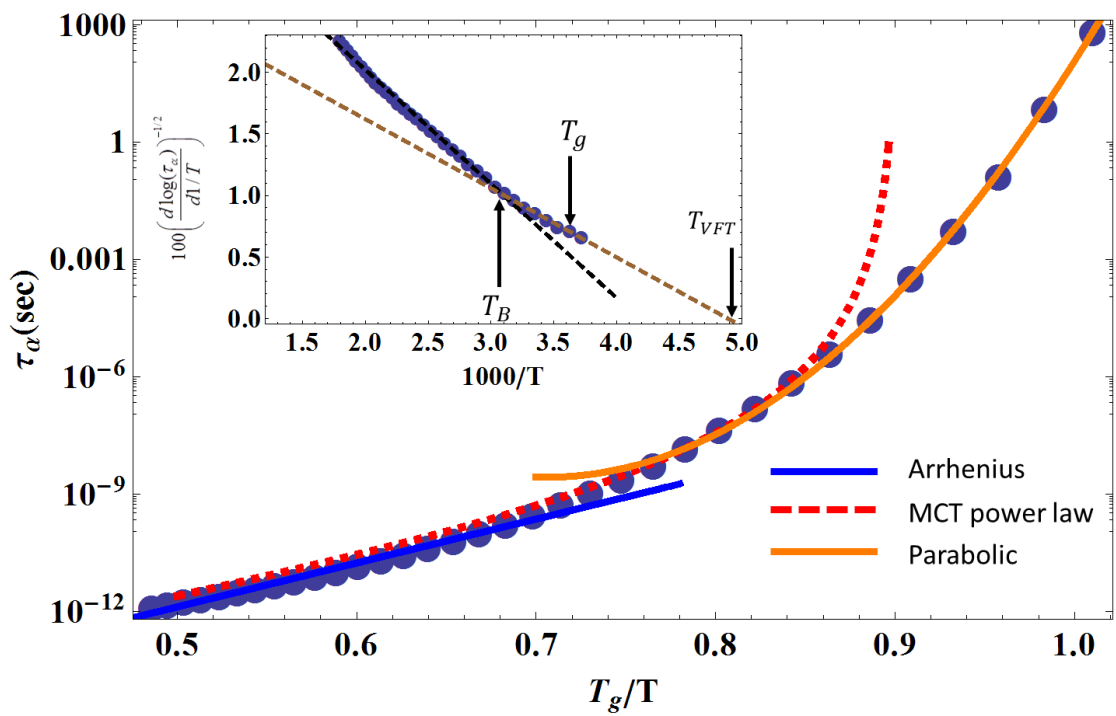

**Figure 4**

## Supporting Information

The alpha relaxation time is often determined in simulation and experiment based on the incoherent dynamic structure factor, $F_s(q,t)$. Here we theoretically analyze this function to develop an expression for $\tau_\alpha$ (Eq. 2 in the text). Using standard Mori-Zwanzig methods[1], the incoherent dynamic structure factor can be written in Laplace space (z) as[2]

$$F_s(q,z) = \left[ z + \frac{q^2}{\Sigma_s(q,z)} \right]^{-1} \tag{1}$$

where the q-dependent memory function is

$$\Sigma_s(q,t) = \beta^2 \left\langle \vec{f}(0) e^{i\vec{q}\bullet\vec{r}(0)} \bullet \vec{f}(t) e^{-i\vec{q}\bullet\vec{r}(t)} \right\rangle \tag{2}$$

and $\vec{f}(t)$ is the total force at time t on a tagged particle at $\vec{r}(t)$. Motivated by the idea that the time correlation of the force decays more rapidly than the phase factors which are directly sensitive to mass transport, the memory function is factorized as

$$\Sigma_s(q,t) \approx \beta^2 \left\langle \vec{f}(0) \bullet \vec{f}(t) \right\rangle \left\langle e^{i\vec{q}\bullet(\vec{r}(t)-\vec{r}(0))} \right\rangle \tag{3}$$

The force time correlation function has a rapidly decaying part due to short time dynamics and a slowing varying part due to hopping-induced structural relaxation[2]:

$$\beta^2 \left\langle \vec{f}(0) \bullet \vec{f}(t) \right\rangle \left\langle e^{i\vec{q}\bullet(\vec{r}(t)-\vec{r}(0))} \right\rangle \rightarrow \beta\zeta_s\delta(t) + \beta^2 \left\langle \vec{f}(0) \bullet \vec{f}(t \rightarrow "\infty") \right\rangle e^{-q^2 r_{loc}^2/6} e^{-2t/\tau_{hop}}$$

$$\simeq \beta\zeta_s\delta(t) + \frac{3}{r_L^2} e^{-q^2 r_{loc}^2/6} e^{-2t/\tau_{hop}} \tag{4}$$

Here we have used the simple MCT relation for the transient localized state that relates the (ideally) arrested part of the force autocorrelation and localization length[3]

$$r_{loc}^2 \left\langle \vec{f}(0) \bullet \vec{f}(t \rightarrow "\infty") \right\rangle = K_0 r_{loc}^2 = 3k_B T \tag{5}$$

a Gaussian Debye-Waller factor to describe the localized state, and modeled memory function relaxation via hopping on the Kramers mean first passage time scale[1,2]

$$\tau_{hop} = \frac{2\pi\tau_s}{\sqrt{K_0 K_B}} e^{\beta(F_B + F_{elastic})} \qquad (6)$$

where $\tau_s = g_d^2 \tau_E$, $\tau_E^{-1} = 8\sqrt{\pi k_B T / 9M} \, \rho d^2 g_d$ is the Enskog binary collision rate, d the sphere diameter, M the particle mass, $\rho$ the fluid number density, and $g_d$ the contact value of the radial distribution function.

A mean structural relaxation time is defined from the z→0 limit of Eq(1) evaluated at the peak of the structure factor, q=q$^*$. Using Eq(5) and $q^* r_{loc} \ll 1$ we obtain

$$\tau_{relax} = \frac{\tau_s}{(q^* d)^2} + \frac{\tau_s \pi}{(q^* d)^2}\sqrt{\frac{K_0}{K_B}} e^{\beta(F_B + F_{elastic})} \equiv \frac{\tau_s}{(q^* d)^2} + \tau_\alpha \qquad (7)$$

The first term reflects the short time process, while the second term is identified as the alpha relaxation time, Eq(2) of the main text. There are *no* adjustable or unknown parameters in this expression, and thus $\tau_\alpha$ can be a priori computed below the predicted dynamic crossover temperature, $T_A$. We emphasize that if $\tau_\alpha$ was identified with Eq(6), none of our conclusions change, and the quantitative aspects are only slightly modified.

*Thermodynamic Input to Theoretical Calculations for Specific Liquids.*

As explained in the main text, experimental dimensionless compressibility data can be well described by the relation $\left(S_{0,site}^{expt}(T)\right)^{-1/2} = -A + \left(B / T\right)$. For OTP and TNB we employ literature values of the density and isothermal compressibility[4] to compute the dimensionless compressibility as a function of temperature, and at 1 atm find it is well described by $A = 0.43, B = 1067 K$. Analogous analysis of glycerol data[5] yields $A = -1.25, B = 990 K$.